\documentstyle[12pt]{report}
\begin{document}
\baselineskip=20pt

\begin{center}
{\bf SCALING AND CROSSOVER IN THE LARGE-N MODEL} \\
{\bf FOR GROWTH KINETICS}\\
{}~~\\
A.Coniglio$^{a}$, P.Ruggiero$^{a}$ and M. Zannetti$^{b}$\\
{}~~\\
$^{a}${\small \it Dipartimento di Scienze Fisiche, Universit\`{a}
di Napoli\\
Mostra d'Oltremare, Padiglione 19, 80125 Napoli, Italy}\\
{}~~\\
$^{b}${\small \it Dipartimento di Fisica, Universit\`{a} di Salerno\\
84081 Baronissi (SA), Italy}
\end{center}
{}~~\\
\begin{center}
{\bf Abstract}
\end{center}

\noindent The dependence of the scaling properties of the
structure factor on space dimensionality, range of interaction,
initial and final conditions, presence or absence of a
conservation law is analysed in the framework of the large-N
model for growth kinetics. The variety of asymptotic
behaviours is quite rich, including standard scaling,
multiscaling and a mixture of the two. The different scaling
properties obtained as the parameters are varied are controlled
by a structure of fixed points with their domains of
attraction. Crossovers
arising from the competition between distinct fixed points
are explicitely obtained.
Temperature fluctuations below the
critical temperature are not found to be irrelevant
when the order parameter is conserved.
The model is solved by integration
of the equation of motion for the structure factor
and by a renormalization group approach.

{}~~\\

05.70.Fh  64.60.Cn  64.60.My  64.75.+g

{}~~\\
{}~~\\

\newpage

\setcounter{chapter}{1}

\section*{1 - Introduction}

\vspace{5mm}

In growth kinetics one deals with the relaxation to equilibrium
of a system quenched from high to low temperature$^{1}$. The processes
of interest are those which exhibit scaling$^{2}$ in the asymptotic
time regime. Denoting with $T_{I}$ and $T_{F}$ the initial and
final temperatures, these processes can be grouped into three
classes characterized by $(T_{I}>T_{c},T_{F}<T_{c})$,
$(T_{I}>T_{c},T_{F}=T_{c})$ and $(T_{I}=T_{c},T_{F}<T_{c})$
where $T_{c}$ is the critical temperature. This subdivision
arises from renormalization group arguments$^{3}$ whereby the
temperature axis (Fig.1) is controlled by three fixed points
at $T=0,T_{c},T=\infty$ and $T_{c}$ is unstable both with
respect to $T=0$ and $T=\infty$. Such a flow diagram leads
naturally to the three universality classes listed above
whose basic processes are those originating and terminating
in a fixed point. By far, the most studied among these is
the phase ordering process from $T_{I}=\infty$ to $T_{F}=0$.

The reason for the continuing interest in this problem
is the persistent lack
of a full understanding of scaling which is observed
both in laboratory$^{4}$ and numerical$^{5}$ experiments. In terms
of the structure factor (Fourier
transform of the equal time order parameter correlation
function) this asymptotic scaling behaviour is of the form
\begin{equation}
C(\vec{k},t) \sim L^{\alpha}(t)F(kL(t))
\end{equation}
where $L(t)$ is a characteristic length which
grows in time with a power law
\begin{equation}
L(t) \sim t^{1/z}.
\end{equation}
A scaling pattern of this type, which we refer to as standard scaling,
is completely characterized by the pair of exponents $z,\alpha$ and
by the scaling function $F(x)$. These quantities depend to a different
extent on the various elements entering in the specification of
the process$^6$ which, in addition to the classes $(T_{I},T_{F})$
discussed above, include the space dimensionality of the
system, the vector dimensionality of the order parameter,
the presence or absence of a conservation law
and the short or long range character of interactions.

The purpose of this paper is to explore in detail the
dependence of the scaling properties on the totality of these
elements in the framework of the large-N model$^{7}$. This,
at the moment, is the only available non trivial soluble model
with a structure sufficiently rich to be adequate for
this kind of investigation. The picture which emerges in the end is quite
informative and exposes clearly the profound
difference between processes with and without conservation
of the order parameter.

We consider a system described by an N-component order parameter
$\vec{\phi}(\vec{x})=(\phi_{1}(\vec{x}),...,\phi_{N}(\vec{x}))$
and by a free energy functional of the Ginzburg-Landau type
\begin{equation}
{\cal H}[\vec{\phi}] =\frac{1}{2} \int d^{d}x \left [
(\nabla \vec{\phi})^{2} +r \vec{\phi}^{2}
+\frac{g}{2N} (\vec{\phi}^{2})^{2} \right ] + {\cal H}_{LR}[\vec{
\phi}]
\end{equation}
where  ${\cal H}_{LR}[\vec{\phi}]$ contains
the long range interaction and will be specified
in section 2. The Gibbs equilibrium states $P_{eq}[\vec{\phi}] \sim
\exp (-\frac{1}{T} {\cal H}[\vec{\phi}])$ are parametrized
by the temperature $T$ and by the pair of coupling constants
$\mu = (r,g)$.
In the large-N limit $(N \rightarrow \infty)$
there is a critical temperature $T_{c}(\mu)
\sim -r/g$ and a phase diagram (Fig.2) in the three
dimansional parameter space $(T,\mu)$ with a surface of critical
points separating ordered states below it from disordered states
above it. The interesting
portion of this phase diagram is the
$(r \leq 0,g \geq 0)$ sector at or below the critical surface
where scaling is to be expected in a quench process.
As we shall see in the following, the set of states at $T=0$
on the $g$-axis plays a special role since it is located at the edge
of both the critical surface and the ordering region
below it.

As anticipated above, the characterization of a
process requires in the order:

\noindent i) specification of the space dimensionality $d$

\noindent ii) specification of the initial condition. This we
do by taking an initial structure factor of the form
\begin{equation}
C(\vec{k},0) = \frac{\Delta}{k^{\theta}}
\end{equation}
where $\Delta$ is a constant and the value of $\theta$
selects the initial state of interest: $\theta=0$
corresponds to an uncorrelated initial state at infinite
temperature $(T_{I}=\infty)$ while $\theta =2$ corresponds to the
critical point $(T_{I}=T_{c})$

\noindent iii) choice between a non conserved order parameter
(NCOP) and a conserved order parameter (COP)

\noindent iv) specification of the short or long range
nature of the interaction

\noindent v) specification of the final state. The interesting
subsets in the equilibrium phase diagram are:

\noindent $[T_{F}=T_{c}=0, \mu_{1}=(r=0,g=0)]$ trivial
critical state at zero temperature

\noindent $[T_{F}=T_{c}>0,\mu_{1}=(r=0,g=0)]$ trivial
critical states at finite temperature ($T$-axis)

\noindent $[T_{F}=T_{c}=0,\mu_{2}=(r=0,g>0)]$ non trivial
critical states at zero temperature ($g$-axis)

\noindent $[T_{F}=T_{c}>0,\mu_{3}=(r<0,g>0)]$ non trivial
critical states at finite temperature (critical surface)

\noindent $[T_{F}<T_{c}, \mu_{3}=(r<0,g>0)]$ phase ordering
region.

\noindent It is convenient to regard the space dimensionality,
the initial condition and the range of the interaction as
forming, so to speak, the environment of the process, while the
set $(T_{F},\mu)$ and the specification NCOP or COP as
elements of discrimination which we will use to identify
processes.

Solving the model analytically and by renormalization
group (RG) we arrive at the following picture.
The asymptotic scaling properties $[z,\alpha ,F(x)]$ depend
on $(T_{F},\mu)$. There is a universality class, under
each heading NCOP or COP, for each of the five regions
$(T_{F},\mu)$ listed above.
In RG language this means that there are five fixed points
$(T_{F}^{\ast},\mu^{\ast})$. The flow in the parameter
space and therefore the extension of the universality
classes depends on the relative stability of these fixed
points. This in turn is regulated by the existence of
critical dimensionalities which depend on the environment,
i.e. initial condition and range of interaction.

The deep difference between NCOP and COP emerges from
how the scaling properties depend on the final state
$(T_{F},\mu)$. The most
striking difference is obtained for quenches inside the
phase ordering region. It was found previously$^{8}$ that
when the system is quenched to $(T_{F}=0,\mu_{3})$
the standard scaling form (1.1) holds only for NCOP, while for
COP it is replaced by the more general multiscaling behaviour
\begin{equation}
C(\vec{k},t) \sim L^{\alpha (x)}(t) F(x)
\end{equation}
where also the exponent $\alpha$ depends on $x=kL(t)$.
We find now that, with some modifications to be
discussed below, this basic distinction NCOP-standard
scaling and COP-multiscaling holds not just for
quenches to $T_{F}=0$,
but for quenches anywhere in the phase ordering  region
$(T_{F}<T_{c},\mu_{3})$. Furthermore, while temperature
perturbations with $0<T_{F}<T_{c}$ are irrelevant for NCOP, it is
not so for COP. For quenches elsewhere, i.e. on the critical
surface, standard scaling holds both for NCOP and COP.
However, while with NCOP $(T_{F},\mu)$ affects $\alpha$ with no
impact on $z$, the opposite occurs for COP.

The analytical tractability of the large-N model$^{7,8,9}$ allows to expose
nicely the mechanism underlying the picture outlined
above and to compute in addition to the asymptotic properties
also the crossovers induced by the competing fixed points.
The question of the extension of the properties of the
large-N model to finite N needs to be treated with care.
We shall comment on this in the concluding section.

The paper is organized as follows: in section 2 the general
features of the large-N model are presented, section 3 is devoted
to the solution of the model by integration of the equation of
motion for the structure factor and in section 4 the model is
analysed by RG methods. Concluding remarks are made in section 5.

\vspace{8mm}

\setcounter{chapter}{2}
\setcounter{equation}{0}

\section*{2 - The large-N model}

\vspace{5mm}

The long range part of the free energy functional (1.3)
is of the form
\begin{equation}
{\cal H}_{LR}[\vec{\phi}] = \int d^{d}x \int d^{d}x^{\prime}
\vec{\phi}(\vec{x}) \cdot V(\vec{x}-\vec{x}^{\prime})
\vec{\phi}(\vec{x}^{\prime})
\end{equation}
with the large distance behaviour $V(\vec{x}-\vec{x}^{\prime})
\sim \mid \vec{x}-\vec{x}^{\prime} \mid ^{-(d+\sigma)}$.
Considering a time evolution of the order parameter governed by
the time dependent Ginzburg-Landau model
and neglecting$^{10,11}$ $k^{2}$ with respect to $k^{\sigma}$
for $\sigma<2$ the equation of motion
for the Fourier transform of the order parameter in the
large-N limit is given by$^{7}$
\begin{equation}
\frac{\partial \phi_{\alpha}(\vec{k},t)}{\partial t} = -\Gamma
\left [ wk^{p+\sigma} + k^{p}R(t) \right ] \phi_{\alpha}(\vec{k},t) +
\eta_{\alpha}(\vec{k},t)
\end{equation}
where $(\alpha=1,...,N)$, $w$ is a coefficient originating in the
small momentum expansion of the interaction,
 $\Gamma$ is a kinetic coefficient,
$p=0$ for NCOP, $p=2$ for COP,
$\:\vec{\eta}(\vec{k},t)\:$ is a gaussian white noise
with expectations
\begin{eqnarray}
<\vec{\eta}(\vec{k},t)> & = & 0 \\
<\eta_{\alpha}(\vec{k},t)\eta_{\beta}(\vec{k}^{\prime},
t^{\prime})> & = & 2\Gamma T_{F} k^{p} \delta_{\alpha
\beta} \delta(\vec{k}+\vec{k}^{\prime}) \delta(t-t^{\prime})
\end{eqnarray}
and
\begin{equation}
R(t) = r +gS(t)
\end{equation}
with $\:S(t) = <\phi_{\alpha}^{2}(\vec{x},t)> \:$ which
is independent of $\:\alpha\:$ and must be determined
self-consistently$^{12}$.
In the following we will let
$\:\sigma\:$ and $\:p\:$ vary continously since
$\:\sigma<2\:$ describes long range interactions while
$\:0<p<2\:$ describes non local conservation of the
order parameter$^{13}$.

Integrating Eq.(2.2) with a random initial condition
$\:\vec{\phi}(\vec{k},0)\:$ and dropping the label
$\alpha$ we find
\begin{equation}
\phi(\vec{k},t) = \phi(\vec{k},0)D(\vec{k},t) + \int_{0}^{t}dt^{\prime}
\eta(\vec{k},t^{\prime})\frac{D(\vec{k},t)}{D(\vec{k},t^{\prime})}
\end{equation}
where
\begin{equation}
D(\vec{k},t) = \exp (- \Gamma  [wk^{p+\sigma}t + k^{p}Q(t)])
\end{equation}
and
\begin{equation}
Q(t) = \int_{0}^{t} dt^{\prime} R(t^{\prime}).
\end{equation}

{}From (2.6) correlation functions of arbitrary order can
be obtained forming products of $\:\phi(\vec{k},t)\:$ and
averaging over both initial condition and thermal noise.
For the average order parameter we find
\begin{equation}
<\phi(\vec{k},t)> = <\phi(\vec{k},0)> D(\vec{k},t)
\end{equation}
which shows that if the initial state is symmetric
$\:<\phi(\vec{k},0)>=0\:$, as we shall assume in the following,
then $\:<\phi(\vec{k},t)>=0\:$
for all time, i.e. dynamics does not brake the symmetry.
The more general case of a time evolution with broken symmetry
is outlined in Appendix I.

The structure factor
$<\phi(\vec{k},t)\phi(\vec{k}^{\prime},t)> =
C(\vec{k},t)\delta(\vec{k}+\vec{k}^{\prime})$
is given by the sum of two contributions
\begin{equation}
C(\vec{k},t) = C_{1}(\vec{k},t) + C_{2}(\vec{k},t)
\end{equation}
where
\begin{equation}
C_{1}(\vec{k},t) = C(\vec{k},t=0)D^{2}(\vec{k},t)
\end{equation}
\begin{equation}
C_{2}(\vec{k},t) = 2\Gamma k^{p}T_{F}D^{2}(\vec{k},t)
 \int_{0}^{t} dt^{\prime}D^{-2}(\vec{k},t^{\prime}).
\end{equation}

{}From (2.10-2.12) one can easily verify that the structure factor
obeys the equation of motion
\begin{equation}
\frac{\partial C(\vec{k},t)}{\partial t} =
-2 \Gamma  [wk^{p+\sigma}+k^{p}R(t)]C(\vec{k},t)+2\Gamma
k^{p}T_{F}
\end{equation}
which is closed by the self-consistency condition
\begin{equation}
S(t)=\int \frac{d^{d}k}{(2\pi)^{d}}C(\vec{k},t).
\end{equation}

Let us now analyse the final equilibrium states. Since there is
no symmetry breaking, it is sufficient to look at the
structure factor in the limit $\:t \rightarrow \infty\:$
obtaining (Appendix II) from Eq. (2.13)
\begin{equation}
C(\vec{k},\infty) = \frac{T_{F}}{wk^{\sigma}+\xi^{-\sigma}} +
(2\pi)^{d}M^{2}\delta (\vec{k})
\end{equation}
where $\:\xi=[R(\infty)]^{-1/\sigma}\:$ is the
equilibrium correlation length. Enforcing the self-consistency
condition (2.14) one finds (Appendix II) that there exists
a critical temperature
\begin{equation}
T_{c} = -\frac{r}{gB(0)}
\end{equation}
such that
\begin{eqnarray}
\left\{ \begin{array}{ll}
\xi^{-1}>0 & \mbox{and $M=0$ for $T_{F}>T_{c}$} \\
\xi^{-1}=0 & \mbox{and  $M^{2}=M_{0}^{2}(T_{c}-T_{F})/T_{c}$
for $T_{F} \leq T_{c}$}
\end{array}
\right.
\end{eqnarray}
with $\:M_{0}^{2}=-r/g\:$ and
\begin{equation}
B(0)= \int \frac{d^{d}k}{(2\pi )^{d}} \frac{1}{wk^{\sigma}}
\propto \frac{1}{d-\sigma}.
\end{equation}

Thus, in the $(T_{F},\mu)$ parameter space (Fig.2)
the critical point as $\:r\:$ and $\:g\:$
are varied ($\:g\:$ must be positive in order to have a well
defined theory) spans a surface which lies on the
$\:r\leq 0\:$ sector and separates ordered states
underneath it, where the structure factor dispalys a
Bragg peak, from the disordered states above it without
Bragg peak. Notice that from (2.16) and (2.18) $\lim_{d \rightarrow
\sigma} T_{c}=0$, implying that $\sigma$ is the
lower critical dimensionality of the model.

As discussed in the Introduction, in the rest of the paper
we shall be concerned with the solution of Eq.(2.13) with
values of $(T_{F},\mu)$ lying on or below the critical
surface.

\vspace{8mm}

\setcounter{chapter}{3}
\setcounter{equation}{0}

\section*{3 - Quench processes}

\vspace{5mm}

{}From (2.10) and (1.4) the formal solution for the structure
factor is given by
\begin{equation}
C(\vec{k},t)=\frac{\Delta}{k^{\theta}} e^{-2\Gamma
[wk^{p+\sigma}t+k^{p}Q(t)]}
+2\Gamma T_{F} k^{p}\int_{0}^{t}dt^{\prime} e^{-2\Gamma
[wk^{p+\sigma}(t-t^{\prime})+k^{p}(Q(t)-Q(t^{\prime}))]}.
\end{equation}
We must now extract the scaling properties.

\vspace{5mm}

\noindent {\bf 3.1 - NCOP}

\vspace{5mm}

With $p=0$ by dimensional analysis we can identify the characteristic
length
\begin{equation}
L(t)=(2\Gamma t)^{1/\sigma}
\end{equation}
which enters (3.1) in the combination $x=kL(t)$. It is
worth to rewrite (3.1) with this change of variable
\begin{equation}
C(\vec{k},t)=\Delta e^{-2\Gamma Q(t)}\frac{L^{\theta}}
{x^{\theta}}e^{-wx^{\sigma}} + \sigma T_{F}L^{\sigma}x^{-\sigma}
\int_{0}^{x} dx^{\prime} x^{\prime \sigma-1}
e^{-w(x^{\sigma}-x^{\prime \sigma})} e^{-2 \Gamma [Q(t)-
Q(t^{\prime})]}
\end{equation}
which shows that (3.2) is the only available choice for
the caracteristic length and therefore necessarily
$z=\sigma$ for any process with NCOP. In order
to solve for $Q(t)$ we integrate (3.3) over $\vec{k}$ and we use
the definitions (2.8) and (2.14) to derive the equation
\begin{equation}
\frac{1}{2\Gamma} \frac{d}{dt}e^{2\Gamma Q(t)} = re^{2\Gamma Q(t)}
+g\Delta A(t) + 2\Gamma T_{F}g\int_{0}^{t}dt^{\prime}A_{0}(t-t^{\prime})
e^{2\Gamma Q(t^{\prime})}
\end{equation}
where
\begin{equation}
A(t)=\int\frac{d^{d}k}{(2\pi)^{d}}\frac{e^{-2\Gamma wk^{\sigma}t}}
{k^{\theta}} = L^{\theta -d}(t) K_{d} \int_{0}^{\Lambda L}
dx x^{d-\theta -1}e^{-wx^{\sigma}}
\end{equation}
with $K_{d}=(2^{d-1} \pi^{d/2} \Gamma(d/2))^{-1}$, $\Lambda$
is a monetum cutoff and $A_{0}(t)$ is the same as $A(t)$ with
$\theta =0$.

\vspace{5mm}

\noindent {\boldmath $T_{F} = 0$}

\vspace{5mm}

Let us now make a further restriction by considering quenches to
zero temperature. With $T_{F}=0$ (3.4) can be integrated obtaining
\begin{equation}
e^{2\Gamma Q(t)}=e^{2\Gamma rt} \left [ 1+2\Gamma g\Delta \left (
\int_{0}^{t_{0}}dt^{\prime} e^{-2\Gamma rt^{\prime}}A(t^{\prime})
+\int_{t_{0}}^{t}dt^{\prime}e^{-2\Gamma rt^{\prime}}A(t^{\prime})
\right ) \right ].
\end{equation}
Next, choosing $t_{0}$ such that $\Lambda L(t_{0}) \sim 1$, from
(3.5) we may approximate
\begin{equation}
L^{d-\theta}(t) A(t) \sim \left\{ \begin{array}{ll}
K_{d}\int_{0}^{\Lambda L}dx x^{d-\theta-1} = K_{d}[\Lambda
L(t)]^{d-\theta}/(d-\theta) & \mbox{for $t<t_{0}$} \\
I=\int_{0}^{\infty}dx x^{d-\theta-1} e^{-wx^{\sigma}}
& \mbox{for $t>t_{0}$}
\end{array}
\right.
\end{equation}
and inserting into (3.6) finally we get
\begin{equation}
e^{2\Gamma Q(t)} =\kappa e^{2\Gamma rt}+g\Delta I e^{2\Gamma rt}
\int_{t_{0}}^{t}dt^{\prime}e^{-2\Gamma rt^{\prime}} L^{\theta-d}(t^{\prime})
\end{equation}
where $\kappa = 1+g\Delta K_{d} \Lambda^{d-\theta}(1-e^{-2\Gamma rt_{0}})/
(d-\theta)r$.

The next step is to extract the behaviour of $\:Q(t)\:$ from
(3.8) after specifying the set of coupling constants $\:\mu=(r,g).
\:$

\vspace{8mm}

\noindent {\boldmath $\:\mu_{1}=(r=0,g=0)\:$}

\vspace{8mm}

\noindent In this case $\:Q(t) \equiv 0 \:$. Thus from (3.3)
we get
\begin{equation}
C(\vec{k},t) = \Delta L^{\theta}F(x)
\end{equation}
with
\begin{equation}
F(x)=\frac{e^{-wx^{\sigma}}}{x^{\theta}}.
\end{equation}
Note that
$\:C(\vec{k},t)\:$ obeys the scaling form (1.1) from beginning
to end over the entire time history of the process with
$\:\alpha = \theta\:$. In this case dynamics propagates
the scaling properties of the initial condition.

\vspace{8mm}

\noindent {\boldmath $\:\mu_{2}=(r=0,g>0)\:$}

\vspace{8mm}

\noindent Setting $\:r=0\:$ in Eq.(3.8)
\begin{equation}
e^{2\Gamma Q(t)} = \kappa + g\frac{\Delta I \sigma}
{\theta +\sigma -d} \left [ L^{\theta+\sigma-d} -
L_{0}^{\theta+\sigma-d} \right ]
\end{equation}
which gives
\begin{equation}
C(\vec{k},t) \sim \frac{L^{\theta}}{\kappa +
\sigma I g \Delta [L^{d_{c}-d}
-L_{0}^{d_{c}-d}]/(d_{c}-d)}F(x)
\end{equation}
with $\:d_{c}=\theta+\sigma\:$, $\:L_{0}=L(t_{0})\:$ and
$F(x)$ given by (3.10). Now
scaling is no more an exact property obeyed over the entire
time history, but only asymptotically. Furthermore, Eq.(3.12)
shows that $\:d_{c}\:$
is a critical dimensionality, in the sense that different
asymptotic behaviours are obtained depending on
$\:d>d_{c}\:$ or $\:d<d_{c}\:$.
For $\:d>d_{c}\:$ the asymptotic behaviour is the same as
for the quench to
$\:\mu_{1}\:$, with the denominator producing a correction
to scaling at early times. Instead for $\:d<d_{c}\:$ there is
a crossover time
$\:t^{\ast} \sim (\Delta g)^{\frac{\sigma}{d-d_{c}}}\:$
such that
\begin{equation}
C(\vec{k},t) \sim \left\{ \begin{array}{ll}
                    L^{\theta}F(x) & \mbox{for $t \ll t^{\ast}$} \\
                    L^{d-\sigma}F(x) & \mbox{for $t \gg t^{\ast}$.}
                   \end{array}
           \right.
\end{equation}
At $\:d=d_{c}\:$ we find
a logarithmic correction due to marginality
\begin{equation}
C(\vec{k},t) \sim \frac{L^{\theta}}{\log L/L_{0}}F(x).
\end{equation}

In order to illustrate the above results
we have integrated numerically the equation of motion (2.13)
for $p=0$, $\sigma=2$, $d=3$, $r=0$, $T_{F}=0$
and different values of $\:g\:$.
Since the crossover involves
only the exponent $\:\alpha\:$ it is convenient to discuss it in
terms of $\:S(t)\:$. The analytical behaviour of $S(t)$ is obtained
integrating (3.12) over $\:\vec{k}\:$
\begin{equation}
S(t) \sim \frac{L^{\theta - d}}{\kappa +\sigma I
g\Delta[L^{d_{c}-d}-L_{0}^{d_{c}-d}]/(d_{c}-d)}.
\end{equation}
The case $\:d>d_{c}\:$ is realized taking
$\:\theta=0\:$, which yields $\:d_{c}=2\:$. The double
logarithmic plot of the numerical solution for
$\:S(t)\:$ shows (Fig.3)
that the power law $\:L^{-d}=t^{-3/2}\:$ is asymptotically
obeyed with deviations from it at early times which become
smaller as $\:g\:$ decreases, i.e. as $\:\mu_{1}\:$ is approached
along the $\:\mu_{2}\:$ axis. Conversely  for $\:d<d_{c}\:$,
obtained by setting $\:\theta=2\:$ which implies $\:d_{c}=4\:$,
the numerical solution for $\:S(t)\:$
behaves as $\:L^{-1} \sim t^{-1/2}\:$
for $\:t \ll t^{\ast}\:$ and as $\:L^{-2} \sim t^{-1}\:$
for $\:t \gg t^{\ast}\:$ (Fig.4).

\vspace{8mm}

\noindent {\boldmath $\:\mu_{3}=(r<0,g>0)\:$}

\vspace{8mm}

\noindent With these values of $r$ and $g$ the
quench is made  inside the region of the coexisting ordered
phases, which is the process normally considered in
the kinetics of phase ordering. In a process of this type
usually one makes
the distinction between the early stage of exponential growth
due to the instability generated by $\:r<0\:$ and the late stage
characterized by scaling behaviour.
However,  due to the existence of crossovers,we should expect
to find  a scaling regime also at {\it early time}
with the exponents appropriate to $\:\mu_{1}\:$ or $\:\mu_{2}\:$
when $\:r\:$ and $\:g\:$ are pushed sufficiently close to
$\:\mu_{1}\:$ or $\:\mu_{2}\:$.
In fact for short time $\:(t<1/2\Gamma |r|)\:$ Eq. (3.8) yields
\begin{equation}
e^{2\Gamma Q(t)} \sim e^{2\Gamma rt} \left ( \kappa  +
\sigma I \frac{g\Delta}{d_{c}-d}[L^{d_{c}-d}-L_{0}^{d_{c}-d}]
\right )
\end{equation}
which implies for the structure factor
\begin{equation}
C(\vec{k},t) \sim \frac{L^{\theta}}{\kappa+
\sigma I g\Delta[L^{d_{c}-
d}-L_{0}^{d_{c}-d}]/(d_{c}-d)} F(x)e^{-2\Gamma rt}
\end{equation}
showing that in front of the exponential actually
there is a prefactor identical to (3.12).
Hence if $|r|$ is sufficiently small one can detect a scaling regime
identical to the one discussed for quenches to $\:\mu_{2}\:$
and {\it preceding} the usual early time regime of
exponential growth.

Instead, when time is large we can set to zero the left
hand side of (3.4) obtaining for the structure factor the
late stage scaling behaviour
\begin{equation}
C(\vec{k},t) \sim M_{0}^{2} L^{d}F(x)
\end{equation}
for any $\:d\:$. This means that for quenches inside the phase
ordering region there is not an  upper critical dimensionality
such that above it one obtains an asymptotic behaviour with
$\alpha=\theta$.

The crossover structure is illustrated in Fig.5 through
the behaviour of $\:S(t)\:$ computed numerically for $\:d=3,
\theta=0,\sigma=2, g=1\:$ and decreasing values of $\:r\:$ in order
to explore the influence on $\:S(t)\:$ of the power law
associated to the $\:\mu_{2}\:$-axis. While away from
$\:\mu_{2}\:$, e.g. at $\:r=-1.0\:$, the behaviour of $\:S(t)\:$
shows only the asymptotic scaling regime $\:S(t) \sim M_{0}^{2}
=1\:$, as $r$ is decreased and $\:\mu_{2}\:$
is approached  for $r$ sufficiently small, e.g. $\:r=-0.01\:$,
$\:S(t)\:$ indeed displays at intermediate times the power
law $\:L^{-d} \sim t^{-3/2}\:$ which we have found previously as an
asymptotic behaviour in the quenches to $\:\mu_{2}\:$.
The case with $\:\theta=2\:$ yields qualitatively similar
results.

\vspace{5mm}

\noindent {\boldmath $T_{F} > 0$}

\vspace{5mm}

For quenches to a finite final temperature $0<T_{F} \leq T_{c}$
we restrict the choice of the initial condition to
high temperature taking $\theta=0$ in (1.4).

\vspace{8mm}

\noindent {\boldmath $\mu_{1}=(r=0,g=0)$}

\vspace{8mm}

\noindent For quenches to the trivial critical states at finite
temperature ($T$-axis in Fig.2), setting $Q(t) \equiv 0$ in (3.3),
we find
\begin{equation}
C(\vec{k},t) = \Delta e^{-wx^{\sigma}}+L^{\sigma}T_{F}F_{0}(x)
\end{equation}
where
\begin{equation}
F_{0}(x)=(1-e^{-wx^{\sigma}})/wx^{\sigma}.
\end{equation}
Thus, for these processes the asymptotic scaling behaviour
is given by $L^{\sigma}T_{F}F_{0}(x)$ while the strength $\Delta$
of initial correlations is
an irrelevant parameter with correction to scaling
behaviour $\sim L^{-\sigma}$.

\vspace{8mm}

\noindent {\boldmath $\mu_{3}=(r<0,g>0)$}

\vspace{8mm}

\noindent The solution of Eq.(3.4) in the general case
has been obtained by Newman and
Bray$^{12}$. For large time and for $\sigma <d <2\sigma$ they find
\begin{equation}
e^{-2\Gamma Q(t)} \sim L^{\omega}(t)
\end{equation}
with
\begin{equation}
\omega = \left\{ \begin{array}{ll}
                  \epsilon=2\sigma-d & \mbox{for $T_{F}=T_{c}$} \\
                  d & \mbox{for $T_{F} < T_{c}.$}
                 \end{array}
         \right.
\end{equation}
Inserting into (3.1) we find
\begin{equation}
C(\vec{k},t) = C(\vec{k},t_{0})\frac{L^{\omega}(t)e^{-wx^{\sigma}}}
{L^{\omega}(t_{0})e^{-wx^{\sigma}_{0}}}
+L^{\sigma}T_{F}\int_{0}^{1-(x_{0}/x)^{\sigma}}dy(1-y)
^{-\omega/\sigma}e^{-wx^{\sigma}y}
\end{equation}
where $t_{0}$ is some microscopic short time such that scaling
holds for $t>t_{0}$ and where $x_{0}=kL(t_{0})$. For $T_{F}<T_{c}$,
recalling that $T_{c}>0$ requires $d>\sigma$ and using (3.22)
the asymptotic scaling behaviour is still given by (3.18) with
$T_{F}$ acting as an irrelevant perturbation
whose correction to scaling behaviour is given by $\sim
L^{\sigma-d}$. For $T_{F}=T_{c}$, the role of
the two terms in the right hand side of (3.23) is reversed.
The thermal contribution provides the dominant scaling term
$\sim L^{\sigma}$ while the first term yields
correction to scaling $\sim L^{\sigma-d}$.
Taking $\Delta=0$ and setting the coupling constant $g$
at the fixed point value$^{14,15}$ the first term can be made to
vanish $(t_{0} \rightarrow 0)$ obtaining
\begin{equation}
C(\vec{k},t) \sim L^{\sigma}T_{c}F_{\epsilon}(x)
\end{equation}
with
\begin{equation}
F_{\epsilon}(x)=\int_{0}^{1}dy(1-y)^{-\epsilon/\sigma}
e^{-wx^{\sigma}y}.
\end{equation}
Notice that $\lim_{\epsilon \rightarrow 0} F_{\epsilon}(x)
= F_{0}(x)$. Thus, at
the upper critical dimensionality $2\sigma$ the non
trivial fixed point merges with the trivial fixed point on
the temperature axis.

In summary, with NCOP we have found the following asymptotic
scaling properties

\noindent 1) $[T_{F}=0,\mu_{1}]$
\begin{equation}
C(\vec{k},t) \sim L^{\theta}(t)F(x)
\end{equation}

\noindent 2) $[T_{F}=T_{c}>0,\mu_{1}]$
\begin{equation}
C(\vec{k},t) \sim L^{\sigma}(t)T_{c}F_{0}(x)
\end{equation}

\noindent 3) $[T_{F}=0,\mu_{2}]$
\begin{equation}
C(\vec{k},t) \sim \left\{ \begin{array}{ll}
L^{\theta}(t) F(x) & \mbox{for $d>d_{c}$} \\
\frac{L^{\theta}(t)}{\log L(t)}F(x) & \mbox{for $d=d_{c}=\theta+\sigma$} \\
L^{d-\sigma}(t) F(x) & \mbox{for $d<d_{c}$}
\end{array}
\right.
\end{equation}

\noindent 4) $[T_{F}=T_{c}>0,\mu_{3}]$
\begin{equation}
C(\vec{k},t) \sim \left\{ \begin{array}{ll}
L^{\sigma}(t) T_{c}F_{0}(x) & \mbox{ for $d \geq 2\sigma$} \\
L^{\sigma}(t) T_{c}F_{\epsilon}(x) & \mbox{for $d<2\sigma$}
\end{array}
\right.
\end{equation}
\noindent 5) $[T_{F}<T_{c}, \mu_{3}]$
\begin{equation}
C(\vec{k},t) \sim L^{d}(t) F(x)
\end{equation}
where $L(t) \sim t^{1/\sigma}$, $x=kL(t)$ and
\begin{eqnarray}
F(x) & = & e^{-wx^{\sigma}}/x^{\theta} \\
F_{0}(x) & = & (1-e^{-wx^{\sigma}})/wx^{\sigma} \\
F_{\epsilon}(x) & = &  \int_{0}^{1} dy(1-y)^{-\epsilon/\sigma}
e^{-wx^{\sigma}y}
\end{eqnarray}
with $\epsilon=2\sigma -d$.

\vspace{8mm}

\noindent {\bf 3.2 - COP}

\vspace{8mm}

\noindent Let us now go back to Eq(3.1) with $p \neq 0$.
The important novelty is that now by dimensional analysis
we can form the two lengths
$L(t) = (2\Gamma t)^{\frac{1}{p+\sigma}}$ and
$\lambda(t) = (2\Gamma \mid Q \mid)^{1/p}$.
In order to establish the scaling behaviour
we must know which of the two is the dominant one.

\vspace{5mm}

\noindent {\boldmath $T_{F} = 0$}

\vspace{5mm}

Considering first, as before, quenches to zero temperature
the structure factor is given by
\begin{equation}
C(\vec{k},t)=\frac{\Delta}{k^{\theta}}e^{-[w(kL)^{p+\sigma}
+a(k\lambda)^{p}]}
\end{equation}
where $a=sign(Q)$. To begin with let us rescale lengths
with respect to $L(t)$
\begin{equation}
C(\vec{k},t) = \Delta e^{-a\beta
x^{p}}L^{\theta}\hat{F}_{>}(x)
\end{equation}
where
\begin{equation}
\beta=\left ( \frac{\lambda}{L} \right )^{p}
\end{equation}
\begin{equation}
\hat{F}_{>}(x)=\frac{e^{-wx^{p+\sigma}}}{x^{\theta}}
\end{equation}
with  $x=kL(t)$ and the reason for the notation $\hat{F}_{>}(x)$
will be clear below.
Again, we must solve for $Q(t)$ using the analogue of Eq.(3.4)
\begin{equation}
\frac{dQ}{dt} = r+g\Delta L^{\theta -d}K_{d}
\int_{0}^{\Lambda L}dxx^{d-\theta -1}
e^{-(wx^{p+\sigma}+a \beta x^{p})}.
\end{equation}

\vspace{8mm}

\noindent {\boldmath $\:\mu_{1} = (r=0,g=0)\:$}

\vspace{8mm}

\noindent Since in this case $\:Q(t) \equiv 0\:$ implying $\lambda(t)
\equiv 0$, clearly $L(t)$ is the dominant length
and $z=p+\sigma$. From (3.35) we have, as in the NCOP case, that standard
scaling is exactly obeyed over the entire time history with
\begin{equation}
C(\vec{k},t) = \Delta L^{\theta}\hat{F}_{>}(x).
\end{equation}

When the nonlinearity is present it is no longer
possible, contrary to NCOP case, to solve directly
for $\:Q(t)\:$. Hence, we now make statements
about the solution of (3.38) for large time by
consistency checks on the assumption that either one of the two
lengths $\:L(t)\:$ and $\:\lambda(t)\:$ prevails over
the other.

\vspace{8mm}

\noindent {\boldmath $\:\mu_{2} = (r=0,g>0)\:$}

\vspace{8mm}

\noindent From (2.5) and (2.8) in this case $\:a\:$ is positive.
Let us first suppose that $L$ prevails
over $\lambda$, i.e. $\:\beta(t) \rightarrow 0\:$ as
$\:t \rightarrow \infty \:$. For large time Eq.(3.38) can
be replaced by
\begin{equation}
\frac{dQ}{dt} = g \Delta L^{\theta -d} K_{d} \int_{0}^{\infty}
dx x^{d-\theta-1}e^{-wx^{p+\sigma}}
\end{equation}
and integrating we find
\begin{equation}
Q(t) \sim \frac{L^{d_{c}+p-d}}{d_{c}+p-d} + const
\end{equation}
where $\:d_{c}=\theta+\sigma\:$. Inserting this
result into (3.36) follows
\begin{equation}
\beta(t) \sim \left\{ \begin{array}{ll}
                      L^{-p} & \mbox{for $d>d_{c}+p$} \\
                      L^{-p}\log L & \mbox{for $d=d_{c}+p$} \\
                      L^{d_{c}-d} & \mbox{for $d<d_{c}+p$}
                      \end{array}
               \right.
\end{equation}
which is consistent with the assumption if $\:d>d_{c}\:$.
Thus, for
$\:d>d_{c}\:$ from (3.35) we find
\begin{equation}
C(\vec{k},t) \sim L^{\theta} \left [ 1-\beta (t) x^{p}
\right ]\hat{F}_{>}(x).
\end{equation}
This result together with (3.42) shows
that there exists yet another critical dimensionality
$\:\tilde{d_{c}}=d_{c}+p\:$ affecting the behaviour of
corrections to scaling.

Assuming next that $\:\beta(t)\:$ takes a constant value $c$,
we find that this is consistent only for $\:d=d_{c}\:$
and from (3.35) follows
\begin{equation}
C(\vec{k},t) \sim L^{\theta}\hat{F}(x)
\end{equation}
with
\begin{equation}
\hat{F}(x)=\frac{1}{x^{\theta}}e^{-[wx^{p+\sigma}+cx^{p}]}
\end{equation}
showing that there is scaling
with the same exponents as at $\:\mu_{1}\:$, but with a modified
scaling function.

Finally, assuming $\:\beta(t) \rightarrow \infty\:$,
from (3.38) follows
\begin{equation}
\beta(t) \sim L^{\frac{p}{(d-\theta+p)}(d_{c}-d)}
\end{equation}
which for consistency requires $\:d<d_{c}\:$. Therefore,
now we must scale with respect to $\lambda$ obtaining
\begin{equation}
C(\vec{k},t) \sim \lambda^{\theta}[1-w\beta ^{-\sigma}(t)x^{\prime
p+\sigma}]\hat{F}_{<}(x^{\prime})
\end{equation}
where
\begin{equation}
\hat{F}_{<}(x^{\prime})=\frac{e^{-x^{\prime p}}}{x^{\prime \theta}}
\end{equation}
and $x^{\prime}=k\lambda(t)$, $\lambda(t) \sim t^{1/z}$
with
\begin{equation}
z=d+p-\theta.
\end{equation}
This result is interesting because a pattern qualitatively
different from the corresponding case with NCOP is obtained.
In the latter case when the nonlinearity becomes relevant
below $d_{c}$ the exponent $\alpha$ changes from the value
$\theta$ to the dimensionality dependent value $(d-\sigma)$,
while the exponent $z$ and the scaling function remain the
same as for the quench to the
trivial fixed point $\mu_{1}$. With COP instead we find that
$\alpha$ keeps always the trivial value $\theta$, while
there is a change in the scaling function and the growth exponent
picks up the value (3.49) dependent on the space dimensionality of the system.

\vspace{8mm}

\noindent {\boldmath $\:\mu_{3} =(r<0,g>0)\:$}

\vspace{8mm}

\noindent For the quench in the phase ordering region
Eq.(3.38) can be rewritten in the
form
\begin{equation}
\frac{dQ}{dt} = r+\frac{g\Delta}{L^{d-\theta}}K_{d}I(\beta)
\end{equation}
where
\begin{equation}
I(\beta) = \int_{0}^{\infty} dx x^{d-\theta-1}e^{-\beta f(x)}
\end{equation}
\begin{equation}
f(x) = ax^{p}+w\beta^{-1} x^{p+\sigma}.
\end{equation}
Making the assumptions to be verified a posteriori that
$\:\beta\:$ is asymptotically divergent and that $a$ is
negative we can make a
saddle point evaluation of (3.51)
\begin{equation}
I(\beta) \sim 2 \left [ \frac{\pi}{w\sigma (p+\sigma)} \right ]
^{\frac{1}{2}}  e^{\frac{w\sigma}{p}u}u^{\gamma}
\end{equation}
where
\begin{equation}
u= \left [\frac{p\beta}{w(p+\sigma)}\right ]^{\frac{p+\sigma}{\sigma}}
\end{equation}
and
\begin{equation}
\gamma = \left [ 2(d-\theta) - (p+\sigma) \right ]/2(p+\sigma).
\end{equation}
For large time we can set to zero the left hand side of (3.50)
obtaining $\:I(\beta)=-rL^{d-\theta}/gK_{d}\Delta\:$. Next, using
(3.53) and taking the logarithm we find the asymptotic relation
\begin{equation}
u=\frac{p(d-\theta)}{w\sigma}\log L - \frac{p\gamma}{w\sigma}
\log u.
\end{equation}
Inserting the leading contribution into (3.54) we obtain
\begin{equation}
\beta(t) \sim (\log L)^{\frac{\sigma}{(p+\sigma)}}
\end{equation}
consistently with the assumption.
Thus, in this case $\:\lambda(t)\:$ and $\:L(t)\:$
diverge in the same way up to a logarithmic factor.
As fundamental length we choose$^{8}$ the inverse
of the wave vector $\:k_{m}(t)\:$ where $\:C(\vec{k},t)\:$
reaches its maximum value
\begin{equation}
k_{m}(t)= L^{-1} u^{\frac{1}{p+\sigma}}.
\end{equation}
Inserting this result in (3.35) and introducing the
variable $\bar{x}=k/k_{m}$ we obtain
\begin{equation}
C(\vec{k},t) \sim \frac{\Delta L^{\theta}}{u^{\theta /
(p+\sigma)}} \frac {e^{u\varphi(\bar{x})}}{\bar{x}^{\theta}}
\end{equation}
where
\begin{equation}
\varphi(\bar{x}) = \frac{p+\sigma}{p}\bar{x}^{p} - \bar{x}^{p+\sigma}.
\end{equation}
Next iterating once (3.56) and inserting the result into (3.59)
we obtain the asymptotic expression$^{8,10}$
\begin{equation}
C(\vec{k},t) \sim \frac{\left [ k_{m}^{\theta-d}(k_{m}L)^
{\frac{p+\sigma}{2}} \right ] ^{\frac{p}{\sigma}\varphi(\bar{x})}}
{(k_{m} \bar{x})^{\theta}}
\end{equation}
which, up to a logarithmic factor in the amplitude, is in the multiscaling
form (1.5)
\begin{equation}
C(\vec{k},t) \sim k_{m}^{-\alpha_{0}(\bar{x})}\frac{1}{\bar{x}^{\theta}}
\end{equation}
with
\begin{equation}
\alpha_{0}(\bar{x})=\frac{(d-\theta)p}{\sigma}\varphi(\bar{x})+\theta.
\end{equation}
This exponent $\alpha_{0}(\bar{x})$ is plotted in Fig.6
for $\theta=0$, $\sigma=2$ showing that the transition from multiscaling
to standard scaling is a smooth one as $p$ is varied continously from
$p=2$ to $p=0$.

Notice that from (3.58) and $k_{m}^{-1} \sim t^{1/z}$, the
exponent $z$ develops a time dependence which asymptotically
is given by
\begin{equation}
z \sim (p+\sigma) \left [ 1+ \frac{\log \log t}{\log t} \right].
\end{equation}

\vspace{5mm}

\noindent {\boldmath $T_{F} > 0$}

\vspace{5mm}

\noindent {\boldmath $\mu_{1}=(r=0,g=0)$}

\vspace{8mm}

\noindent Again, with $Q(t)\equiv 0$ and $\theta=0$
it is straightforward to derive from (3.1)
the standard scaling result
\begin{equation}
C(\vec{k},t)= \Delta e^{-wx^{p+\sigma}} +L^{\sigma}T_{F}
\hat{F}_{0}(x)
\end{equation}
where
\begin{equation}
\hat{F}_{0}(x)=(1-e^{-wx^{p+\sigma}})/wx^{\sigma}
\end{equation}
and the same considerations made about (3.19) apply here.

\vspace{8mm}

\noindent {\boldmath $\mu_{3}=(r<0,g>0)$}

\vspace{8mm}

For quenches to $\mu_{3}=(r<0,g>0)$ now we
cannot solve explicitely for $Q(t)$. Then we proceed differently
by rewriting (3.1) as
\begin{equation}
C(\vec{k},t) = \Delta e^{u(t)\varphi(\bar{x})}+2\Gamma T_{F}
k^{p} \int_{0}^{t}dt^{\prime}e^{u(t)\varphi(\bar{x})-u(t^{\prime})
\varphi(\bar{x}^{\prime})}
\end{equation}
where $\varphi(\bar{x})$ has
been defined in (3.60) and $u(t)$ is related to
$Q(t)$ by (3.54). In the following we
will drop the bar over $x$. If by analogy with what we
have found at $T_{F}=0$ we make the ansatz
\begin{equation}
e^{u(t)}=cL^{\rho}(t)
\end{equation}
with $\rho>0$, the quantities to be determined are the constant
$c$ and the exponent $\rho$. Inserting into (3.67), and
assuming $k_{m}^{-1}(t) \sim L(t)$ up to logarithmic
factors we find
\begin{equation}
C(\vec{k},t)=\Delta (cL^{\rho})^{\varphi(x)} \left [
1+T_{F}L^{\sigma}\frac{1}{wx^{\sigma}}\int_{0}^{x} dx^{\prime}
x^{\prime p+\sigma -1} (x/x^{\prime})^{\rho \varphi(x^{\prime})}
e^{-\rho \varphi(x^{\prime})\log L} \right ].
\end{equation}
For large $L$ the integral in the right hand side can be evaluated
by steepest descent. Keeping in mind that $\rho>0$ and that
$\varphi(x)$ behaves as in Fig.6, we must distinguish between
$x<x^{\ast}$ and $x>x^{\ast}$ where $x^{\ast}=(\frac{p+\sigma}
{p})^{1/\sigma}$ is the non trivial zero of $\varphi(x)$.
For $x<x^{\ast}$ the exponential reaches the maximum value at
$x^{\prime}=0$, while for $x>x^{\ast}$ the maximum is at
$x^{\prime}=x$ yielding
\begin{equation}
C(\vec{k},t) \sim  \Delta(cL^{\rho})^{\varphi(x)}+
\frac{T_{F}L^{\sigma}}{wx^{\sigma}}(cL^{\rho})^{\varphi(x)}
\end{equation}
for $x<x^{\ast}$ and
\begin{equation}
C(\vec{k},t) \sim \Delta (cL^{\rho})^{\varphi(x)} +
\frac{T_{F}L^{\sigma}}{wx^{\sigma}}
\end{equation}
fro $x>x^{\ast}$.

We now determine $c$ and $\rho$ by using the self-consistency
condition (2.14). From (3.54), where $\beta$ is defined in (3.36),
and (3.68) follows $R(t) = \dot{Q}(t) \sim (\rho \log L)^{
\frac{\sigma}{p+\sigma}} L^{-\sigma}$. On the other hand $R(t)$ can
be computed from (2.5) with $S(t)$ obtained by
integration of  (3.70) and (3.71) over $\vec{k}$.
Thus, for $T_{F}=0$ dropping logarithmic factors one finds
\begin{equation}
L^{-\sigma} = r+g\Delta K_{d}
L^{-d}\int_{0}^{\Lambda L}dx x^{d-1} e^{\varphi(x)\log (cL^{\rho})}.
\end{equation}
For large $L$ the integral picks up the dominant contribution at
$x=1$ giving
\begin{equation}
L^{-\sigma}=r+g\Delta K_{d} c^{\sigma/p}L^{(\rho\sigma -pd)/p}.
\end{equation}
The right hand side can vanish only if $\rho=pd/\sigma$,
in agreement with (3.63), and
$c^{\sigma/p} \sim M_{0}^{2}.$

For $T_{F} \neq 0$, the first terms in the right hand side of
(3.70) and (3.71) are asymptotically negligible with respect to
the second ones yielding in place of (3.72)
\begin{equation}
L^{-\sigma} = r+gT_{F}L^{\sigma-d}K_{d} \left [ \int_{0}^{x^{\ast}}
dx x^{d-\sigma-1}(cL^{\rho})^{\varphi(x)} + \int_{x{\ast}}
^{\Lambda L}dx x^{d-\sigma-1} \right ].
\end{equation}
Furthermore, keeping into account that
for large $L$ the dominant contribution to the first integral is
obtained at $x=1$ and that the contribution at $x^{\ast}$ is
negligeble in the second one finally we obtain
\begin{equation}
L^{-\sigma}=r(\frac{T_{F}-T_{c}}{T_{c}})+gT_{F}K_{d}c^{\sigma/p}
L^{\rho\frac{\sigma}{p}+\sigma-d}.
\end{equation}
For $T_{F}<T_{c}$ this implies $\rho=p(d-\sigma)/\sigma$ and
$c^{\sigma/p} \sim M_{0}^{2}(\frac{T_{c}-T_{F}}{T_{F}T_{c}})$.
Inserting into (3.70) and (3.71) one finds
\begin{equation}
C(\vec{k},t) \sim T_{F}\frac{L^{\alpha(x)}}{wx^{\sigma}}
\end{equation}
with
\begin{equation}
\alpha(x)= \left \{ \begin{array}{ll}
\frac{p}{\sigma}(d-\sigma)\varphi(x)+\sigma & \mbox{for
$x<x^{\ast}$} \\
\sigma & \mbox{for $x>x^{\ast}$}
\end{array}
\right.
\end{equation}
showing that the structure factor obeys multiscaling
for $x<x^{\ast}$ and standard scaling for $x>x^{\ast}$.
Notice that the two behaviours match at $x^{\ast}$ since
$\varphi(x^{\ast})=0$.

If the quench is made on the critical surface $(T_{F}=T_{c})$,
then Eq. (3.75) gives $\rho=\frac{p}{\sigma}(d-2\sigma)$
which is negative below the upper critical dimensionality
contradicting the initial assumption $\rho>0$. Hence we take $\rho=0$
which implies standard scaling with $u(t)$ constant.
{}From (3.68) and (3.36) we have $Q(t)=bL^{p}(t)$.
Inserting  in (3.1) and taking $\Delta=0$ for simplicity, we obtain
\begin{equation}
C(\vec{k},t)=L^{\sigma} T_{c} \hat{F}_{\epsilon}(x)
\end{equation}
with
\begin{equation}
\hat{F}_{\epsilon}(x)=\frac{1}{wx^{\sigma}}\int_{0}^{x}dx^{\prime}
x^{\prime p+\sigma-1}e^{-[w(x^{p+\sigma}-x^{\prime p+\sigma})
+b(x^{p}-x^{\prime p})]}.
\end{equation}
{}From the self-consistency condition (2.14) now we find
\begin{equation}
b=\frac{p+\sigma}{2\Gamma p} g K_{d}L^{2\sigma-d}T_{c}
\left \{ \int_{0}^{\infty} dx x^{d-1} [\hat{F}_{\epsilon}
(x) -\frac{1}{wx^{\sigma}}] -\int_{\Lambda L}^{\infty}
[\hat{F}_{\epsilon}(x) -\frac{1}{wx^{\sigma}}] \right \}
\end{equation}
where we have used the definition (2.16) of $T_{c}$. From
(3.79) for large $x$ we have
$\hat{F}_{\epsilon} \sim \frac{1}{wx^{\sigma}}(1-cbx^{-\sigma})$
where $c=\frac{p}{p+\sigma}\int_{0}^{\infty}d\psi \psi e^{-\psi}$.
Inserting into (3.80) and defining $f(b)=\int_{0}^{\infty} dx
x^{d-1} [\hat{F}_{\epsilon}(x) -\frac{1}{wx^{\sigma}}]$ we have
\begin{equation}
b \left [ 1-\frac{(p+\sigma)gK_{d}T_{c}\Lambda^{d-2\sigma}}
{2\Gamma p (d-2\sigma)} \right ] =\frac{p+\sigma}{2\Gamma p}
gK_{d}L^{2\sigma-d}f(b)
\end{equation}
from which follows $b=0$ for $d>2\sigma$, while for $d<2\sigma$
the value of $b$ is given by the condition $f(b)=0$.
Solving this equation in the $\epsilon$-expansion$^{15}$
one finds $b \sim \epsilon$ as for NCOP,
which implies the same structure
of fixed points. Namely, for $\epsilon \rightarrow 0$
the non trivial fixed point still merges with the trivial one
and $\lim_{\epsilon \rightarrow 0} \hat{F}_{\epsilon}(x)=\hat{F}_{0}(x)$.

The summary of the asymptotic properties with COP is given hereafter

\noindent 1) $[T_{F}=0,\mu_{1}]$
\begin{equation}
C(\vec{k},t) \sim L^{\theta}(t) \hat{F}(x)
\end{equation}

\noindent 2) $[T_{F}=T_{c},\mu_{1}]$
\begin{equation}
C(\vec{k},t) \sim L^{\sigma}(t)T_{c} \hat{F}_{0}(x)
\end{equation}

\noindent 3) $[T_{F}=0,\mu_{2}]$
\begin{equation}
C(\vec{k},t) \sim \left\{ \begin{array}{ll}
L^{\theta}(t) \hat{F}_{>}(x) & \mbox{for $d>d_{c}$} \\
L^{\theta}(t)\hat{F}(x) & \mbox{for $d=d_{c}=\theta+\sigma$} \\
\lambda^{\theta}(t) \hat{F}_{<}(x^{\prime})
& \mbox{for $d<d_{c}$}
\end{array}
\right.
\end{equation}
where $x=kL(t)$ and $x^{\prime}=k\lambda(t)$

\noindent 4) $[T_{F}=T_{c}>0,\mu_{3}]$
\begin{equation}
C(\vec{k},t) \sim \left\{ \begin{array}{ll}
L^{\sigma}(t) T_{c}\hat{F}_{0}(x) & \mbox{for $d \geq 2\sigma$} \\
L^{\sigma}(t) T_{c}\hat{F}_{\epsilon}(x) & \mbox{for $d<2\sigma$}
\end{array}
\right.
\end{equation}

\noindent 5) $[0<T_{F}<T_{c},\mu_{3}]$
\begin{equation}
C(\vec{k},t)=T_{F}\frac{L^{\alpha(x)}(t)}{wx^{\sigma}}
\end{equation}

\noindent 6) $[T_{F}=0,\mu_{3}]$
\begin{equation}
C(\vec{k},t) \sim L^{\alpha_{0}(x)}(t)
\frac{1}{x^{\theta}}
\end{equation}
with
\begin{eqnarray}
& & L(t) \sim t^{1/(p+\sigma)} \\
& & \lambda(t) \sim t^{1/(d+p-\theta )} \\
& & \hat{F}_{>}(x)  =  \frac{e^{-wx^{p+\sigma}}}{x^{\theta}} \\
& & \hat{F}(x)=\frac{1}{x^{\theta}}e^{-[wx^{p+\sigma}+cx^{p}]} \\
& & \hat{F}_{<}(x^{\prime})=\frac{e^{-x^{\prime p}}}{x^{\prime \theta}} \\
& &\hat{F}_{0}(x)  =\frac{1}{wx^{\sigma}} (1-e^{-wx^{p+\sigma}}) \\
& & \hat{F}_{\epsilon}(x)=\frac{1}{wx^{\sigma}}\int_{0}^{x}dx^{\prime}
x^{\prime p+\sigma-1}e^{-[w(x^{p+\sigma}-x^{\prime p+\sigma})
+b(x^{p}-x^{\prime p})]} \\
& & \alpha_{0}(x)  = \frac{(d-\theta)}{\sigma}\varphi(x)+\theta.
\end{eqnarray}
\begin{equation}
\alpha(x) = \left\{ \begin{array}{ll}
2+(d-2)\varphi(x) & \mbox{for $x<x^{\ast}$} \\
2 & \mbox{for $x>x^{\ast}$}
\end{array}
\right.
\end{equation}
\begin{equation}
\varphi(x)=\frac{p+\sigma}{p} x^{p}-x^{p+\sigma}.
\end{equation}

\vspace{8mm}

\setcounter{chapter}{4}
\setcounter{equation}{0}

\section*{4 - Renormalization group}

\vspace{5mm}

We now discuss the scaling properties
of the model within the RG
approach to the problem$^{16}$. Let us recall that in static critical
phenomena the Wilson RG equations are obtained performing the
following operations on the equilibrium probability
distribution $\:P_{eq}[\vec{\phi};T_{F},\mu]\:$

\noindent i) elimination of hard modes $\:\vec{\phi}(\vec{k})\:$
with $\:\Lambda/l <k \leq \Lambda\:$ where
$\:l>1\:$

\noindent ii) rescaling of wave vectors and order parameter
$$
\begin{array}{l}
\vec{k}^{\prime}=l\vec{k}\\
\vec{\phi}^{\prime}(\vec{k}^{\prime})=l^{-y}\vec{\phi}(\vec{k})
\end{array}
$$

\noindent iii) requirement of form invariance of
$\:P_{eq}[\vec{\phi};T_{F},\mu]\:$.

\noindent These operations generate recursion relations for
the parameters $\:(T_{F},\mu)\:$ which allow to describe
scaling in terms of the geometry of the fixed
points and their domains of attraction.

In quench processes one deals with a time dependent
probability distribution
$\:P[\vec{\phi};t,T_{F},\mu]\:$. Therefore, RG
transformations performed on this object are expected to give recursion
relations for $\:(t,T_{F},\mu)\:$ with a fixed point structure
which accounts for the variety of scaling behaviours
obtained in section 3.
In order to implement the procedure outlined above, we should
construct $\:P[\vec{\phi};t,T_{F},\mu]\:$ and then
carry out renormalization. However, since the stochastic process
is gaussian and all the equal time information is in
$\:C(\vec{k},t)\:$, we can work directly with the equation
of motion (2.13). First we separate
soft and hard modes
\begin{equation}
C(\vec{k},t)=C_{s}(\vec{k},t)+C_{h}(\vec{k},t)
\end{equation}
with
\begin{equation}
C_{s}(\vec{k},t)=  \left  \{ \begin{array}{ll}
                             C(\vec{k},t) & \mbox{for $0\leq k\leq
\Lambda/l$}\\
                             0 & \mbox{for $\Lambda/l<k\leq \Lambda$}
                             \end{array}
                    \right.
\end{equation}
\begin{equation}
C_{h}(\vec{k},t)=  \left \{  \begin{array}{ll}
                             0 & \mbox{for $0\leq k \leq \Lambda/l$}\\
                             C(\vec{k},t)& \mbox{for $\Lambda/l<k\leq
\Lambda.$}
                             \end{array}
                    \right.
\end{equation}
The equation of motion for either component is given by
\begin{equation}
\frac{\partial C_{s,h}(\vec{k},t)}{\partial t} = -2\Gamma [w k^{p+\sigma}
+k^{p}R(t)]C_{s,h}(\vec{k},t)+2\Gamma k^{p}T_{F}
\end{equation}
with
\begin{equation}
R(t)=r+gS_{s}(t)+gS_{h}(t)
\end{equation}
\begin{equation}
S_{s,h}(t)=\int \frac{d^{d}k}{(2\pi)^{d}}C_{s,h}(\vec{k},t).
\end{equation}
Then we proceed to eliminate hard modes. Integrating the equation
of motion for $\:C_{h}(\vec{k},t)\:$ we find
\begin{equation}
C_{h}(\vec{k},t)=C_{h}(\vec{k},0)e^{-2\Gamma [wk^{p+\sigma}t+
k^{p}Q(t)]}\\
+ 2\Gamma k^{p}T_{F} \int_{0}^{t}dt^{\prime}
e^{-2\Gamma [wk^{p+\sigma}(t-t^{\prime})+k^{p}(Q(t)-Q(t^{\prime}))]}
\end{equation}
where $\:Q(t)=Q_{s}(t)+Q_{h}(t)\:$.
After integrating over $\:\vec{k}\:$ we should solve for
$\:S_{h}(t)\:$ and insert the result into the equation for
$\:C_{s}(\vec{k},t)\:$. However, considering that we must
eliminate modes with $\:k>L^{-1}(t)\:$ and that in the scaling
regime these have already equilibrated, with a good approximation we
can set
\begin{equation}
C_{h}(\vec{k},t) \sim C_{h}(\vec{k},\infty)= \frac{T_{F}}
{wk^{\sigma}+R(\infty)}
\end{equation}
and we are left with Eq.(4.4) for $\:C_{s}(\vec{k},t)\:$ with
\begin{equation}
R(t)=r+gS_{s}(t)+gS_{h}(\infty).
\end{equation}
Next we carry out the rescalings for $k < \Lambda /l$
\begin{equation}
C^{\prime}(\vec{k}^{\prime},0)=l^{d-2y(0)}C(\vec{k},0)
\end{equation}
\begin{equation}
\vec{k}^{\prime}=l\vec{k}
\end{equation}
\begin{equation}
t^{\prime}=l^{-z}t
\end{equation}
\begin{equation}
C^{\prime}(\vec{k}^{\prime},t^{\prime})=l^{d-2y(x)}
C(\vec{k},t)
\end{equation}
where we have allowed for an order parameter scaling index
dependent on the invariant quantity $\:x=kt^{1/z}\:$. From (4.4) and (4.13)
we obtain the transformed equation of motion
\begin{eqnarray}
& & \frac{\partial C^{\prime}(\vec{k}^{\prime},t^{\prime})}
{\partial t^{\prime}}  =  -2\Gamma \left [ l^{z-\sigma -p}
w k^{\prime p+\sigma} +
l ^{z-p}k^{\prime p}R(l^{z}t^{\prime})
+ \Gamma^{-1} \frac{dy(x)}{dt^{\prime}}\log l
\right ]  C^{\prime}(\vec{k}^{\prime}, t^{\prime})
\nonumber \\
& &  +2\Gamma k^{\prime p}l^{z-p+d-2y(x)}T_{F}
\end{eqnarray}
which can be rewritten in the same form as the original
equation of motion (2.13)
\begin{equation}
\frac{\partial C^{\prime}(\vec{k}^{\prime},t^{\prime})}
{\partial t^{\prime}} = -2\Gamma [w^{\prime}k^{\prime p+\sigma}
+k^{\prime p}R^{\prime}(t^{\prime})]
C^{\prime}(\vec{k}^{\prime},t^{\prime})+2\Gamma
k^{\prime p}T_{F}^{\prime}
\end{equation}
defining
\begin{equation}
w^{\prime}=l^{z-\sigma-p}w
\end{equation}
\begin{equation}
R^{\prime}(t^{\prime})=l^{z-p}R(l^{z}t^{\prime})+
\frac{x\frac{dy}{dx}\log l}{\Gamma zt^{\prime}}
\end{equation}
\begin{equation}
T_{F}^{\prime}=l^{z-p+d-2y(x)}T_{F}.
\end{equation}
In order to preserve the self-consistent structure we must require
$$
R^{\prime}(t^{\prime})=r^{\prime}+g^{\prime}S^{\prime}(t^{\prime})
$$
$$
S^{\prime}(t^{\prime}) = \int \frac{d^{d}k^{\prime}}
{(2\pi)^{d}} C^{\prime}(\vec{k}^{\prime},t^{\prime}).
$$
Since the left hand side of (4.17) depends only on $\:t^{\prime}\:$,
the $x$-dependence on the right hand side must disappear
implying
\begin{equation}
y(x)=c\log x+y(0).
\end{equation}
This form of $y(x)$ diverges at $x=0$ and we must necessarily
have $c=0$.

So far we have managed to map the process described by (2.13) in the
new process governed by (4.15) with new parameters
$\Delta^{\prime},w^{\prime},\mu^{\prime},T^{\prime}_{F}$.
We are going to be interested in those processes (fixed points)
whose parameters $\Delta^{\ast},w^{\ast},\mu^{\ast},
T^{\ast}_{F}$ do not change under renormalization.
For these values of the parameters, since the form of the original
equation of motion (4.15) and the self-consistent structure have
been preserved we have $C^{\prime}(\vec{k}^{\prime},t^{\prime})
=C(\vec{k},^{\prime},t^{\prime})$.
The existence of such  fixed points is
compatible only with the standard scaling choice
$y(x) \equiv y$, which seems to exclude the multiscaling
solution found in the previous section. We shall comment
on this later on.

We proceed to extract recursion relations.
{}From (1.4) and (4.10)
follows
\begin{equation}
C^{\prime}(\vec{k}^{\prime},0)=l^{d+\theta-2y}
\Delta/k^{\prime \theta}
\end{equation}
implying
\begin{equation}
\Delta^{\prime}=l^{d+\theta-2y}\Delta.
\end{equation}
Next, rewrite (4.17) as
\begin{equation}
r^{\prime}+g^{\prime}S^{\prime}(t^{\prime})=l^{z-p}[ r+
gS_{s}(l^{z}t^{\prime})+gS_{h}(\infty)]
\end{equation}
where for quenches to final states with $\:R(\infty)=0\:$
one has
\begin{equation}
S_{h}(\infty)=T_{F}B(0)(1-l^{\sigma-d})
\end{equation}
and $\:B(0)\:$ is defined in (2.18).
Since from (4.13) we have
\begin{equation}
S^{\prime}(t^{\prime})=l^{2(d-y)}S_{s}(t)
\end{equation}
inserting (4.23) and (4.24) in (4.22)
we find $\:r^{\prime}=l^{z-p}[r+gT_{F}B(0)(1-l^{\sigma-d})]\:$
and $\:g^{\prime}=l^{z-p+2(y-d)}g\:$.
Introducing the scaling field
\begin{equation}
\tau = r+gT_{F}B(0)= r(T_{c}-T_{F})/T_{c}
\end{equation}
the whole set of recursion relations is given by
\begin{equation}
\Delta^{\prime}=l^{d+\theta-2y}\Delta
\end{equation}
\begin{equation}
w^{\prime}=l^{z-\sigma-p}w
\end{equation}
\begin{equation}
\tau^{\prime}=l^{z-p}\tau
\end{equation}
\begin{equation}
g^{\prime}=l^{z-p+2(y-d)}g
\end{equation}
\begin{equation}
T_{F}^{\prime}=l^{z-p+d-2y}T_{F}.
\end{equation}
We emphasize that the use of the same scaling index $y$ for $t>0$
as well as for $t=0$ leads to fixed
points as  processes for which scaling holds over the entire
time history. For processes where scaling invariance is only
asymptotic is not necessary that there exists a non trivial
fixed point solution of (4.26-4.30).

\vspace{8mm}

\noindent {\bf \it Fixed points}

\vspace{8mm}

\noindent The next step is to look for fixed points
$\:(\Delta^{\ast},w^{\ast},\tau^{\ast},g^{\ast},
T_{F}^{\ast})\:$ of the recursion relations and to extract
from them the exponents $z$ and $\alpha$. These exponents are
determined by imposing that two of the parameters have a
finite fixed point value. In the remaining recursion relations we
consider the trivial fixed point solution and the corresponding
domain of attraction. In order to show how this works in
practice let us begin by imposing that $\Delta^{\ast}$
and $w^{\ast}$ be finite. From this, using (4.26) and (4.27) follows
\begin{equation}
z=p+\sigma
\end{equation}
and
\begin{equation}
\alpha=2y-d=\theta.
\end{equation}
Inserting these values into (4.28),(4.29)and (4.30) we obtain
\begin{equation}
\tau^{\prime}=l^{\sigma}\tau
\end{equation}
\begin{equation}
g^{\prime}=l^{d_{c}-d}g
\end{equation}
\begin{equation}
T_{F}^{\prime}=l^{\sigma-\theta}T_{F}
\end{equation}
with $d_{c}=\theta+\sigma$. The trivial solution is $\tau^{\ast}=
g^{\ast}=T_{F}^{\ast}=0$ which coincides with $(T_{F}=0,\mu_{1})$.
The corresponding domain of attraction, considering that for quenches
from high temperature to the critical surface $\theta \leq
\sigma$, is given by the $g$-axis, i.e. $(T_{F}=0,
\mu_{2})$ for $d>d_{c}$. Otherwise, for$d<d_{c}$, this fixed
point is completely unstable, and (4.31),(4.32) do not apply
for quenches to $(T_{F}=0,\mu_{2})$. This coincides with what
we have found in section 3.

Next, let us impose that $\Delta^{\ast}$ and $g^{\ast}$
be finite. Then we find $z=d+p-\theta$, $\alpha=\theta$ and the
remaining recursion relations
\begin{equation}
w^{\prime}=l^{d-d_{c}}w
\end{equation}
\begin{equation}
\tau^{\prime}=l^{d-\theta}\tau
\end{equation}
\begin{equation}
T_{F}^{\prime}=l^{d-2\theta}T_{F}
\end{equation}
with the solution $w^{\ast}=\tau^{\ast}=T_{F}^{\ast}=0$.
$w$ flows to zero for $d<d_{c}$. This fixed point for
COP corresponds to the quench to $(T_{F}=0,\mu_{2})$.
For NCOP this corresponds to the case of independent
particles and it is possible to show that in this case
the amplitude of the structure factor vanishes.
Thus in order to treat quenches to $(T_{F}=0,\mu_{2})$ for NCOP
with $d<d_{c}$ we must require that $w^{\ast}$ and $g^{\ast}$ be
simultaneously finite. This reproduces immediately the results
$z=\sigma$ and $\alpha=d-\sigma$ of (3.13). The ensuing
recursion relations for the other parameters
\begin{equation}
\Delta^{\prime}=l^{d_{c}-d}\Delta
\end{equation}
\begin{equation}
\tau^{\prime}=l^{\sigma}\tau
\end{equation}
\begin{equation}
T_{F}^{\prime}=l^{2\sigma-d}T_{F}.
\end{equation}
yield a trivial solution which is unstable under all perturbations.
The meaning of $\Delta$ flowing to infinity can be understood
from the result of section 3 where the crossover time $t^{\ast}$
vanishes for $\Delta \rightarrow \infty$. In this limit we
have a fixed point in the sense specified above that the same scaling
behaviour applies over the whole history of the process.

So far we have dealt with fixed points with $T_{F}^{\ast}=0$
and $\tau^{\ast}=0$, namely with quenches to zero temperature
critical points. In order to analyse quenches on the critical surface
at finite temperature, we must require $T_{F}^{\ast}$ and
$w^{\ast}$ finite as it is usually done in
static critical phenomena. From these conditions follows
$z=p+\sigma$, $\alpha=\sigma$ and
\begin{equation}
\Delta^{\prime}=l^{\theta-\sigma}\Delta
\end{equation}
\begin{equation}
\tau^{\prime}=l^{\sigma}\tau
\end{equation}
\begin{equation}
g^{\prime}=l^{2\sigma-d}g.
\end{equation}
Thus, for quenches to finite temperature on the critical
surface $\Delta$ is irrelevant and the attractive fixed point
goes from trivial to non trivial as the dimensionality goes from above
to below the upper critical dimensionality $2\sigma$.

Finally let us come to the discussion of quenches inside the
phase ordering region. For this we require that $w^{\ast}$ and
the fixed point ratio
\begin{equation}
\left ( \frac{\tau}{g} \right ) ^{\ast}=\left [ M_{0}^{2}
(\frac{T_{F}-T_{c}}{T_{c}}) \right ]^{\ast}
\end{equation}
be finite obtaining $z=p+\sigma$ and $\alpha=d$. The other
recursion relations
\begin{equation}
\Delta^{\prime}=l^{\theta-d}\Delta
\end{equation}
\begin{equation}
T_{F}^{\prime}=l^{\sigma-d}T_{F}
\end{equation}
show that temperature perturbations
in the ordering region are irrelevant and that $\Delta$ flows to
zero. Now, with $\Delta^{\ast}=T_{F}^{\ast}=0$ the structure
factor $C(\vec{k},t)$ vanishes identically, namely the scaling
form (3.18) is obeyed with $F(x) \equiv 0$. This means that a
non trivial scaling solution for a quench in the phase ordering
region with $\alpha=d$ is necessarily asymptotic and cannot be
made to hold over the entire history of the process by any
choice of the parameters.

This completes the analysis of the fixed point structure of the
phase diagram and the derivation of exponents.
The RG treatment of the problem presented above reproduces the
whole structure found in section 3 except for multiscaling
in the quench below $T_{c}$ with $p \neq 0$.
The point is that the RG procedure we have followed
above yields the exponents $z$ and $\alpha$
within a standard scaling framework, but gives no
information on the scaling function. If one goes further by
performing the scaling ansatz in the equation of motion
an equation for $F(x)$ is obtained and it turns
out that the scaling function vanishes if $T_{F}<T_{c}$
and $p \neq 0$.
In order to recover multiscaling
trough the RG approach the set of transformations
must be properly generalized. We do this only for
the $(T_{F}=0,\mu_{3})$ case$^{17}$.

As we have seen in section 3, when there is multiscaling $z$
is weakly time dependent.
Let us then generalize the set of transformations
(4.10-4.13) by allowing $z$ to depend on $t$ with the constraint
\begin{equation}
\lim_{t \rightarrow \infty} z(t)=z_{\infty}=p+\sigma.
\end{equation}
With these modifications and $T_{F}=0$ in place of (4.14) we find
\begin{eqnarray}
& & \frac{\partial C^{\prime}(\vec{k}^{\prime},t^{\prime})}
{\partial t^{\prime}}  =  -2\Gamma  [wk^{\prime p+\sigma}
+k^{\prime p}l^{\sigma}R(t)] C^{\prime}(\vec{k}^{\prime},
t^{\prime}) \nonumber \\
& & -2 \left \{ \Gamma (z-z_{\infty})[wk^{\prime p+\sigma}
+k^{\prime p} l^{\sigma}R(t)]+\frac{dy(x)}{dt^{\prime}} \right \}
C^{\prime}(\vec{k}^{\prime},t^{\prime})\log l.
\end{eqnarray}
Imposing the requirement of form invariance we find that (4.49)
is of the form (4.15) if in place of (4.17)
the following conditions are satisfied
\begin{equation}
R^{\prime}(t^{\prime})=l^{\sigma} R(t)
\end{equation}
\begin{equation}
\frac{dy}{dt^{\prime}}=-\Gamma [z(t^{\prime})-z_{\infty}]
[wk^{\prime p+\sigma}+k^{\prime p}R^{\prime}(t^{\prime})].
\end{equation}
This latter equation holds
also for unprimed variables and using
\begin{equation}
\frac{dx}{dt} = \frac{x}{zt} \left [ 1-\frac{t}{z}\frac{dz}{dt}
\log t \right ]
\end{equation}
we obtain
\begin{equation}
\frac{dy}{dx} = -\frac{\Gamma z[z-z_{\infty}]t^{1-(p+\sigma)/z}}
{[1-\frac{t}{z} \frac{dz}{dt} \log t]} \left ( wx^{p+\sigma-1}
+x^{p-1}t^{\sigma/z}R(t) \right ).
\end{equation}
In order to get rid of the time dependence on the right
hand side we must have
\begin{equation}
-\Gamma z(z-z_{\infty}) = ct^{(z-z_{\infty})/z}
(1-\frac{t}{z}\frac{dz}{dt} \log t)
\end{equation}
\begin{equation}
t^{\sigma/z}R(t)=b
\end{equation}
where $b$ and $c$ are constants. Integrating (4.53) we find
\begin{equation}
y(x)=c \left [ \frac{wx^{p+\sigma}}{p+\sigma}+\frac{b}{p}
x^{p} \right ] + y(0).
\end{equation}
{}From (4.55) and
the definition (2.5) of $\:R(t)\:$ follows that the
sign of $b$ is determined by the parameters $\:\mu=
(r,g)\:$. In particular $b$ is a negative quantity at $\mu_{3}$.
Imposing $\:y_{max}=d\:$,
as appropriate for quenches to $\:\mu_{3}\:$,
taking the position of the maximum at $\:x=1\:$
and using $y(0)=(d+\theta)/2$ from (4.56)
we find $\:b=-w\:$, $\:c=(\theta-d)p(p+\sigma)/2\sigma\:$ and
\begin{equation}
2y(x)-d=\alpha_{0}(x)
\end{equation}
where $\:\alpha_{0}(x)\:$ coincides with (3.63). Hence, for quenches to
$\:\mu_{3}\:$, by allowing for a time dependence in the
growth exponent $\:z\:$, we have recovered via RG the multiscaling
behaviour of the exact analytical solution.

The explicit time dependence of $\:z(t)\:$ is obtained
extracting the asymptotic behaviour from (4.54)
\begin{equation}
z(t) \simeq z_{\infty} \left [ 1 + \frac{\log \log t}
{\log t} \right ]
\end{equation}
which is consistent with the assumption (4.48) and reproduces (3.64).
Finally, inserting this result into (4.50) we find
\begin{equation}
R(t) \sim \left ( \frac{\log t}{t} \right )^{\frac{\sigma}{p+\sigma}}
\end{equation}
in agreement with (3.57).

\vspace{8mm}

\setcounter{chapter}{5}
\setcounter{equation}{0}

\section*{5 - Conclusions}

\vspace{5mm}

In this paper we have investigated in detail the solution of
the large-N model for growth kinetics with the aim of giving
a comprehensive view of the influence on the scaling properties
of the various elements which enter in the specification of
the problem. These are the presence or the absence
of a constraint on the order parameter (COP or NCOP), the
initial condition, the structure of
the phase diagram of final equilibrium states, the range
of the interaction and the dimensionality of space.

What the model shows, apart for quenches to the trivial
fixed point, is that scaling properties are quite different
with and without conservation law. This is due to the existence
of only one divergent length $L(t)$ for NCOP and of
two divergent lengths $L(t)$ and $\lambda(t)$ for COP.
It is the interplay between these two lengths which leads to
phenomena not observed with NCOP such as i) a change in the growth
law when crossing the critical dimensionality for the quenches to
$(T_{F}=0,\mu_{2})$ and ii) multiscaling for
quenches inside the phase ordering region.
About multiscaling, the availability of the
rich variety of cases illustrated in the paper should allow to
speculate about its origin. Thus we have found that for NCOP
in no circumstance there is multiscaling and the same holds true
for COP, except for the quenches below $T_{c}$. What, then, is
the peculiarity of these processes. One possible interpretation
is that in these processes the system orders and likes to do so
 by condensing, i.e. by growing a peak which scales like
$L^{d}$, at $\vec{k}=0$. This is fine with
NCOP, but with COP there is a conflict with the conservation law
which prevents anything to happen at $\vec{k}=0$. In this
case the peak is formed at some $\vec{k}_{m} \neq 0$. The
compromise realized in the large-N model is multiscaling
whereby the behaviour $L^{\alpha_{0}(x)}$ of the structure factor
interpolates smoothly between the behaviours $L^{0}$ at $\vec{k}=0$
and $L^{d}$ at $\vec{k}_{m}$. This picture fits nicely with the
absence of multiscaling in any process with NCOP and in all
processes with COP on the critical surface. In fact, in the
latter case there is nothing to condense at $\vec{k}=0$ and
in the former there are no constraints at $\vec{k}=0$.
However, with this mechanism for multiscaling there should be
nothing special about $N=\infty$ and multiscaling should be
found also for $N < \infty$. Numerical simulations$^{18}$
so far have reported no evidence for multiscaling for
$N=1$ and $N=2$ in two and three dimensions.
This could mean that multiscaling
disappears for $N \leq d$, when localized topological
defects appear in the system. However an argument
against the existance of multiscaling for any finite $N$ is the
result of Bray and Humayun$^{19}$. By analysing an equation
of motion for the structure factor which includes the first
order correction in $1/N$, they have reached the conclusion
that multiscaling does not survive for $N<\infty$ since the
correction term sustains a standard scaling solution.

What we can say is that
both standard scaling and multiscaling imply scale invariance
of the structure factor due to the presence of a divergent
length. So far we have not found a criterion to predict
a priori which one should hold. Only a direct calculation,
either analytical or numerical can discriminate between
the two. It must be emphasised that these
concepts apply also to other models. For example in
DLA (diffusion limited aggregation) model in
two dimensions it has been found numerically$^{20}$ that
multiscaling holds. In any case, even if there is not a
general criterion, multiscaling should be more likely to
occur in situations where the width of the interface becomes
very large.

Let us then comment on those features
of the $N=\infty$ solution which we believe to be
of general validity. The crossover structure which
emerges as the parameters of the quench are moved over the
manifold of final equilibrium states is a generic feature
which is expected to hold beyond the large-N model. The main point of our
analysis is that it is quite possible, before the true
asymptotic behaviour is reached, to detect a preasymptotic scaling
behaviour due
to a less stable fixed point lying in the neighbourhood
of the final equilibrium state. It is natural to pose the
question of the observability of these effects.
Here we suggest (Appendix I) that the duration of this
preasymptotic behaviour can be magnified and observed in off-critical
quenches$^{21}$.

Furthermore, the crossover picture we have illustrated
suggests the possibility
of observing a crossover in the growth law (1.2)
in the symmetrical quench of a system with scalar $(N=1)$ COP.
In that case asymptotically
$L(t)$ grows according to (1.2) with $z=3$. On the other hand
in the trivial theory $(r=0,g=0)$ one has $z=4$ for COP,
irrespective of the order parameter being a scalar or a vector.
Thus, for a quench to $(T_{F}=0,\mu_{3})$ sufficiently
close to $(T_{F}=0,\mu_{1})$ with $N=1$ it should be possible to
observe the influence of the trivial fixed point at early time
producing a crossover in (1.2) from $z=4$ to $z=3$.

\vspace{8mm}

\setcounter{chapter}{6}
\setcounter{equation}{0}

\section*{6-Appendix I}

\vspace{5mm}

Let us consider a process where symmetry breaking along one direction
is allowed, e.g.
\begin{equation}
<\phi_{\alpha}(\vec{x},t)>=N^{1/2}M(t)\delta_{\alpha,1}.
\end{equation}
This may be due to non symmetrical initial conditions, or to
the presence of an external field or to both of these circumstances$^{22}$.
Introducing the external field along the $1$-direction
from (1.3) dropping the long range term we obtain the
equation of motion for the order parameter
\begin{eqnarray}
& & \frac{\partial \phi_{\alpha}(\vec{x},t)}{\partial t}=
-\Gamma(i\nabla)^{p}
\left [ \left ( -\nabla^{2}+r+\frac{g}{N}\sum_{\beta=1}{N}
\phi_{\beta}^{2}(\vec{x},t) \right ) \phi_{\alpha}(\vec{x},t) -
h_{\alpha} \right ]
\nonumber \\
& & +\eta_{\alpha}(\vec{x},t)
\end{eqnarray}
where $\:h_{\alpha}=N^{1/2}h\delta_{\alpha,1}\:$.

Defining the fluctuation field $\:\psi(\vec{x},t)\:$ by
\begin{equation}
\phi_{1}(\vec{x},t)=N^{1/2}M(t)+\psi(\vec{x},t)
\end{equation}
and inserting into (6.2) we obtain the pair of equations
\begin{eqnarray}
& & \frac{\partial (N^{1/2}M+\psi)}{\partial t}  =  -\Gamma (i\nabla)^{p}
\left [ -\nabla^{2} \psi +r(N^{1/2}M+\psi) + \right.  \nonumber \\
& & \frac{g}{N} (N^{1/2}M+\psi)\sum_{\beta \neq 1} \phi_{\beta}^{2} +
g(N^{1/2}M^{3}+3M^{2}\psi +\frac{3}{N^{1/2}}M\psi^{2}
+\frac{1}{N}\psi^{3}) - \nonumber \\
& &  \left. N^{1/2}h \right ] +\eta_{1}(\vec{x},t)
\end{eqnarray}
\begin{eqnarray}
& & \frac{\partial \phi_{\beta}}{\partial t}  =  -\Gamma(i\nabla)^{p}
\left [ -\nabla^{2} \phi_{\beta} +r\phi_{\beta}+
\frac{g}{N}\sum_{\gamma  \neq 1}
\phi_{\gamma}^{2}\phi_{\beta} +
g(M^{2}+\frac{2}{N^{1/2}}M\psi \right. \nonumber \\
& & \left. +\frac{1}{N}\psi^{2})\phi_{\beta} \right ] + \eta_{\beta}.
\end{eqnarray}
Taking the large-N limit we replace $\:\frac{1}{N}\sum_{\beta \neq
1}\phi_{\beta}^{2}(\vec{x},t)\:$ by $\:<\phi_{\beta}^{2}(\vec{x},t)>
=S_{\perp}(t)\:$ and collecting terms of the same order of
magnitude, from (6.4) and (6.5) we obtain the set of equations
\begin{equation}
\frac{\partial M(t)}{\partial t}= -\Gamma(i\nabla)^{p} \left (
[r+gM^{2}(t)]M(t)+gS_{\perp}(t)M(t)-h \right )
\end{equation}
\begin{equation}
\frac{\partial \psi(\vec{x},t)}{\partial t}= -\Gamma(i\nabla)^{p}
\left [ -\nabla^{2} +r+3gM^{2}(t)+gS_{\perp}(t) \right ]
\psi(\vec{x},t)+\eta_{1}(\vec{x},t)
\end{equation}
\begin{equation}
\frac{\partial \phi_{\beta}(\vec{x},t)}{\partial t}=-\Gamma(i\nabla
)^{p}
\left [ - \nabla^{2} +r+gM^{2}(t)+gS_{\perp}(t) \right ]
\phi_{\beta}(\vec{x},t) +\eta_{\beta}(\vec{x},t).
\end{equation}
Notice that the behaviour of $\:\psi(\vec{x},t)\:$ is immediately
obtained once the pair of coupled equations (6.6) and (6.8)
have been solved. Furthermore, for COP Eq.(6.6) is trivial
since $\:M(t)\:$ does not change in time and keeps the initial value
$\:M(0)\:$. In this case the equation (6.8) for the
transverse components, after Fourier transforming can be rewritten as
\begin{equation}
\frac{\partial \phi_{\beta}(\vec{k},t)}{\partial t} =
-\Gamma k^{p} \left [-\nabla^{2}+\tilde{r}+gS_{\perp}(t) \right ]
\phi_{\beta}(\vec{x},t)+\eta_{\beta}(\vec{x},t)
\end{equation}
where
\begin{equation}
\tilde{r}=r+gM(0)^{2}.
\end{equation}
Thus $\tilde{r}$ can be modulated by varying $M(0)$.  In particular a
quench to $\mu_{2}$ can be realized as an off-critical quench
to $T_{F}=0$ and at the edge of the coexistence region
$M^{2}(0) = -r/g$.

\vspace{8mm}

\setcounter{chapter}{7}
\setcounter{equation}{0}

\section*{7-Appendix II}

\vspace{5mm}

In the limit $\:t \rightarrow \infty\:$ the left hand side of
Eq.(2.13) vanishes and we have
\begin{equation}
0=-2\Gamma [wk^{p+\sigma}+k^{p}R]C(\vec{k},\infty)+2\Gamma k^{p}
T_{F}
\end{equation}
where $R$ stands for $R(\infty)$.

Let us first consider $\:p=0\:$ and a system in a finite volume
$\:V\:$. Assuming $\:R>0\:$ from (7.1) follows
\begin{equation}
C(\vec{k},\infty)=\frac{T_{F}}{wk^{\sigma}+R}
\end{equation}
where $\:R\:$ must satisfy the self-consistency condition
\begin{equation}
R=r+g\frac{1}{V}\sum_{\vec{k}}\frac{T_{F}}{wk^{\sigma}+R}
\end{equation}
which always admits a solution with $\:R>0\:$.

In the infinite volume limit Eq.(7.3) can be rewritten as
\begin{equation}
R=r+gT_{F}B(R)+g\frac{T_{F}}{VR}
\end{equation}
where, allowing for the possibility that the solution $\:R
\rightarrow 0\:$,
the zero wave vector term in the summation has been separated out
and
\begin{equation}
B(R)=\int \frac{d^{d}k}{(2\pi)^{d}}\frac{1}{wk^{\sigma}+R}
\end{equation}
is a non-negative monotonously decreasing function of $\:R\:$ with
a maximum at $\:B(0)=K_{d}\Lambda^{d-\sigma}/[w(d-\sigma)]\:$.
$\Lambda$ is a momentum cutoff  and $K_{d}=(2^{d-1}\pi^{d/2}\Gamma(d/2))^{-1}$.
{}From Eq.(7.4) follows that when $\:r<0\:$ there is
a critical  temperature
\begin{equation}
T_{c}=-\frac{r}{gB(0)}
\end{equation}
such that for $\:T_{F}>T_{c}\:$ there exists a solution with
$\:R>0\:$ and therefore the last term in the right hand side
can be neglected, but for $\:T_{F}<T_{c}\:$ Eq.(7.4) can
only have the solution $\:R=0\:$ provided that $\:R\:$ vanishes
in the infinite volume limit as $\:R \sim 1/V\:$. Hence, defining
the constant
\begin{equation}
M^{2}=\frac{T_{F}}{VR}
\end{equation}
and inserting into (7.4) we find
\begin{equation}
M^{2}=M_{0}^{2}(T_{c}-T_{F})/T_{c}
\end{equation}
with $\:M_{0}^{2}=-r/g\:$.
Notice that for $\:T_{F}=T_{c}\:$ Eq.(7.4) admits the
solution $\:R=0\:$ provided $\:R \sim V^{-x}\:$ with
$\:0<x<1\:$. In conclusion the structure factor is given by
\begin{equation}
C(\vec{k},\infty) = \left \{ \begin{array}{ll}
T_{F}/(wk^{\sigma}+R) & \mbox{with $R>0$ for $T_{F}>T_{c}$} \\
T_{F}/wk^{\sigma} + (2\pi)^{d}M^{2}\delta(\vec{k}) & \mbox{for
$T_{F}\leq T_{c}$}.
\end{array}
\right.
\end{equation}

For $p \neq 0$ Eq.(7.2) applies only for $\vec{k} \neq 0$.
Due to the conservation law $C(\vec{k}=0,\infty)$ is
determined by the initial condition. If we consider an
initial state without symmetry breaking we have
$\phi(\vec{k}=0)=0$ and $C(\vec{k}=0,t=0)=0$. Then Eq.(7.3)
must be replaced by
\begin{equation}
R=r+\frac{g}{V}\sum_{\vec{k}\neq 0}\frac{T_{F}}{wk^{\sigma}+R}.
\end{equation}
Solving the above equation for $R$ we find that
there exists a temperature
\begin{equation}
\tilde{T}(V)=-\frac{r}{g}\frac{1}{V\sum_{\vec{k}\neq 0}wk^{\sigma}}
\end{equation}
such that  $R \geq 0$ for $T_{F} \geq
\tilde{T}(V)$, while $R <0$  for $T_{F}<\tilde{T}(V)$
with $|R| < k_{min}$ where $k_{min} \sim
V^{-1/d}$ is the minimum value of the wave vector in the summation.
When the infinite volume limit is taken from (7.11) follows
$\tilde{T}(V) \rightarrow T_{c}$ and
the anlogue of (7.4) is
\begin{equation}
R=r+gT_{F}B(R)+\frac{g}{V}\frac{T_{F}}{wk_{min}^{\sigma}+R}.
\end{equation}
For $T_{F}>T_{c}$ there is a solution $R>0$ and the last term
vanishes, while for $T_{F}<T_{c}$ Eq.(7.12) admits the solution
$R=0$ provided $(k_{min}^{\sigma}+R) \sim 1/V$. Writing
\begin{equation}
M^{2}=\frac{T_{F}}{V(wk_{min}^{\sigma}+R)}
\end{equation}
in place  of (7.7), in the end we recover the results
(7.8) and (7.9).

\newpage

{\bf Figure captions}

\vspace{5mm}

\noindent Fig.1 - Renormalization group flow on the temperature axis.

\noindent Fig.2 - Manifold of final equilibrium states with
the critical surface separating disordered states
(above) from ordered states (below).

\noindent Fig.3 - Behaviour of $S(t)$ in a quench to
$\mu_{2}$ for NCOP with $\sigma=2,\Delta=10$ and $d>d_{c}$: $(d=3,
\theta=0)$. The straight dashed lines have slope
$-3/2$.

\noindent Fig.4 - Behaviour of $S(t)$ in a quench to
$\mu_{2}$ for NCOP with $\sigma=2, \Delta=10$ and $d<d_{c}$:
$(d=3,\theta=2)$. The top dashed line has
slope $-0.5$, the others below have slope $-1.0$.

\noindent Fig.5 - Behaviour of $S(t)$ in a quench to
$\mu_{3}$ for NCOP with $\sigma=2, \Delta=10, d=3,\theta=0$
at fixed $g=1$ as $r$ approaches the
$\mu_{2}$-axis. The straight dashed line has slope
$-1.45$.

\noindent Fig.6 - Plot of $\alpha_{0}(x)$ for different values
of $p$. Multiscaling goes over smoothly to standard scaling
as $p \rightarrow 0$.

\newpage

{}~~\\
{}~~\\
{\bf  References}
{}~~\\
\begin{enumerate}

\item Reviews can be found in J.D.Gunton, M. San Miguel
and P.S.Sahni in {\it Phase transitions and critical phenomena},
edited by C.Domb and J.L.Lebowitz (Academic Press, New York,
1983), Vol. 8, p.267; H.Furukawa, {\it Adv. Phys.} {\bf 34},
703, (1985); K.Binder, {\it Rep. Progr. Phys.} {\bf 50}, 783,
(1987)

\item K.Binder and D.Stauffer, {\it Phys. Rev. Lett.} {\bf 33},
1006, (1974);
J.Marro, J.L.Lebowitz and M.H.Kalos, {\it ibid.} {\bf 43},
282, (1979); H.Furukawa, {\it Progr. Theor. Phys.}
{\bf 59}, 1072, (1978); {\it Phys. Rev. Lett.} {\bf 43},
136, (1979)

\item A.J.Bray, Lectures given at NATO Advanced Study Institute on
{\it "Phase Transitions and Relaxation in Systems with
Competing Energy Scales"}, 1993 Geilo, Norway.

\item Y.C.Chou and W.I.Goldburg, {\it Phys. Rev. A} {\bf 23},
858, (1981); N.C.Wong and C.M.Knobler, {\it Phys. Rev. A}
{\bf 24}, 3205, (1981); A.Craievich and J.M.Sanchez,
{\it Phys. Rev. Lett.} {\bf 47}, 1308, (1981);
M.Hennion, D.Rouzaud and P.Guyot, {\it Acta Metall.}
{\bf 30}, 599, (1982); S.Katano and M.Iizumi, {\it
Phys. Rev. Lett} {\bf 52}, 835, (1984);
S.Komura, K.Osamura, H.Fujii and
T.Takeda {\it Phys. Rev. B} {\bf 31}, 1278, (1985);
B.D.Gaulin, S.Spooner and Y.Mori, {\it Phys. Rev. Lett.}
{\bf 59}, 668, (1987)

\item O.T.Valls and G.F.Mazenko, {\it Phys. Rev. B} {\bf 34},
7941, (1986); Y.Oono and S.Puri {\it Phys. Rev. Lett.}
{\bf 58}, 836, (1987) and {\it Phys. Rev. A} {\bf 38}, 434,
1542, (1988);
J.Amar, F.Sullivan and R.Mountain
{\it Phys. Rev. B} {\bf 37}, 196, (1988); T.M.Rogers. K.R.Elder
and R.C.Desai {\it Phys. Rev. B} {\bf 37}, 9638, (1988);
R.Toral, A.Chakrabarti and J.D.Gunton {\it Phys. Rev. B}
{\bf 39}, 4386, (1989); C.Roland and M.Grant {\it Phys.
Rev. B} {\bf 39}, 11971, (1989); M.Mondello and N.Goldenfeld,
{\it Phys. Rev. A} {\bf 42}, 5865, (1990) and {\it Phys. Rev. E}
{\bf 47}, 2384, (1993); A.Shinozaki and
Y.Oono, {\it Phys. Rev. Lett.} {\bf 66}, 173, (1991) and
{\it Phys. Rev. E} {\bf 48}, 2622, (1993)

\item The problem of the definition of universality classes
for growth kinetics as been addressed in
Z.W.Lai, G.F.Mazenko and O.T.Valls {\it Phys. Rev. B} {\bf 37},
9481, (1988)

\item G.F.Mazenko and M.Zannetti,
{\it Phys. Rev. B} {\bf32}, 4565, (1985)

\item A.Coniglio and M.Zannetti, {\it Europhys. Lett.} {\bf 10},
575, (1989)

\item Z.R\'{a}cz and T.T\'{e}l, {\it Phys. Lett.} {\bf 60A},
3, (1977); G.F.Mazenko and M.Zannetti , {\it Phys. Rev. Lett}
{\bf 32}, 4565, (1984);
F.de Pasquale and P.Tartaglia, {\it Phys. Rev. B} {\bf 33},
2081, (1986); A.Coniglio and M.Zannetti, in {\it From phase
transitions to cahos} edited by G.Gyorgyi, I.Kondor, L.Sasv\'{a}ri
and T.T\'{e}l, World Scientific (1992), p. 100

\item H.Hayakawa, Z.R\'{a}cz and T.Tsuzuki. {\it Phys. Rev. E}
{\bf 47}, 1499, (1993)

\item A.J.Bray, {\it Phys. Rev. E} {\bf 47}, 3191, (1993)

\item For a systematic treatment of the $1/N$ expansion see
T.J.Newman and A.J.Bray {\it J. Phys. A: Math. Gen.} {\bf 23}
4491, (1990)

\item P.Tamayo and W.Klein, {\it Phys. Rev. Lett.} {\bf 63},
2757, (1989); A.J.Bray, {\it Phys. Rev. Lett.} {\bf 66},
2048, (1991)

\item H.K.Janssen, B.Schaub and B.Schmittmann,
{\it Z. Phys.} {\bf 73}, 539, (1989)

\item A.J.Bray, {\it Critical exponent and scaling functions
for nonequilibrium critical dynamics} Preprint 1989

\item For real space RG combined with numerical simulation
methods see G.F.Mazenko, O.T.Valls and F.Zhang, {\it Phys.
Rev. B} {\bf 31}, 4453, (1985); Z.W.Lai, G.F.Mazenko and
O.T.Valls, {\it ibid.} {\bf 37}, 9481, (1988); J.Vi\~{n}als,
M.Grant, M.San Miguel, J.D.Gunton and E.T.Gawlinski,
{\it Phys. Rev. Lett.} {\bf 54}, 1264, (1985); S.Kumar,
J.Vi\~{n}als and J.D.Gunton, {\it Phys. Rev. B} {\bf 34},
1908, (1986); C.Roland and M.Grant, {\it Phys. Rev. Lett.}
{\bf 60}, 2657, (1988) and {\it Phys. Rev. B} {\bf 39},
11971, (1989).

\noindent For Wilson momentum shell RG see A.J.Bray,
{\it Phys. Rev. Lett.} {\bf 62}, 2841, (1989) and
{\it Phys. Rev. B} {\bf 41}, 6724, (1990); A.Coniglio and M.Zannetti,
{\it Phys. Rev. B} {\bf 42}, 6973, (1990)

\item A.Coniglio and M.Zannetti, {\it Phys. Rev. B} {\bf 42},
6973, (1990)

\item  A.Coniglio, Y.Oono, A.Shinozaki and M.Zannetti,
{\it Europhys. Lett.} {\bf 18}, 59, (1992); M.Siegert and
M.Rao, {\it Phys. Rev. Lett.} {\bf 70}, 1956, (1993);
M.Mondello and  N.Goldenfeld, {\it Phys. Rev. E} {\bf 47},
2384, (1993)

\item A.J.Bray and K.Humayun, {\it Phys. Rev. Lett.} {\bf 68},
1559, (1992)

\item C.Amitrano, A.Coniglio, P.Meakin and M.Zannetti,
{\it Phys. Rev. B} {\bf 44}, 4974, (1991)

\item A.Coniglio, F.Corberi and M.Zannetti, to be published

\item A.J.Bray and J.G.Kissner, {\it J. Phys. A: Math.
Gen.} {\bf 25}, 31, (1992)
M.Zannetti, {\it J. Phys. A: Math. Gen.} {\bf 26},
3037, (1993)

\end{enumerate}

\end{document}